\documentclass[fleqn,usenatbib]{mnras}

\usepackage{newtxtext,newtxmath}

\usepackage[T1]{fontenc}
\usepackage{ae,aecompl}

\usepackage{graphicx}	
\usepackage{amsmath}	
\usepackage{amssymb}	
\usepackage{syntonly}
\usepackage{trace}
\tracingall

\title[Supernovae generated high Velocity Compact Clouds]{Supernovae generated High Velocity Compact Clouds}

\author[A. Yalinewich and P. Beniamini]{
Almog Yalinewich,$^{1}$\thanks{E-mail: almog.yalin@gmail.com}
Paz Beniamini$^{2,3}$
\\
$^{1}$Canadian Institute for Theoretical Astrophysics, 60 St. George St., Toronto, ON M5S 3H8, Canada\\
$^{2}$Department of Physics, The George Washington University, Washington, DC 20052, USA\\
$^3$Astronomy, Physics and Statistics Institute of Sciences (APSIS)}

\date{Accepted XXX. Received YYY; in original form ZZZ}

\pubyear{2015}

\begin{document}
\label{firstpage}
\pagerange{\pageref{firstpage}--\pageref{lastpage}}
\maketitle

\begin{abstract}
It has been proposed that an intermediate mass black hole can produce compact molecular clouds with large velocity dispersions.
This model is called into question due to the discovery of several such short lived clouds, which suggests they are common, and hence it is unlikely that their formation involves an exotic object such as an intermediate mass black hole.
In this paper we propose an alternative model, where supernovae produce such clouds.
We apply our model to a such a cloud called CO-0.40-0.22.
A previous work pointed out the fact that the kinetic energy of this cloud is comparable to that from a supernova, 
but disqualified the supernova scenario due to lack of thermal emission.
We explain the lack of thermal emission as a consequence of the large density of the medium. In such a medium, the supernova shock is radiation dominated, and hence the initial temperatures are lower by many orders of magnitude compared with a typical supernova. Finally, we discuss some of the consequences of having a significant population of intermediate mass black holes in the galactic centre.
\end{abstract}




\section{Introduction}

Recent observations revealed a number of peculiar molecular clouds, also known as high-velocity compact clouds (HVCCs), in the galactic bulge \citep{tokuyama_oka_et_al_2017,oka_white_et_al_1999}. Their properties are summarised in table \ref{tab:hvcc_params}. The typical projected size of these clouds is less than 10 pc, and their spread in radial velocities is a few tens km/s, significantly larger than expected from their motion in the Galaxy or from their self gravity. One particular cloud, CO-0.40-0.22, rose to prominence as evidence for an intermediate mass black hole \citep{oka_mizuno_et_al_2016}. It has been shown that the cloud's velocity dispersion can result from a scattering event by a compact object with a mass of $10^5 M_{\odot}$ \citep{oka_tsujimoto_et_al_2017}.

Alternative explanations for the observed features have also been considered. One candidate is a supernova \citep{oka_tsujimoto_et_al_2017}. While the kinetic energy of CO-0.40-0.22, $10^{49.7}$ erg, is similar to that of a typical supernova (as are the energies of the other clouds, see table \ref{tab:hvcc_params}), this scenario has been disqualified because no counterpart in other wavelengths has been observed, whereas a typical supernova remnant is expected to emit in the X-ray range \citep{cioffi_et_al_1988}.

Without strong evidence to the contrary, it is reasonable to assume that the observed high-velocity compact clouds share a common origin. Given that there are at least 7 HVCCs, with typical lifetimes (inferred from their sizes and velocities) of $10^5$ years, suggests a formation rate of such clouds of at least $10^{-4} \mbox{yrs}^{-1}$. This alone implies that the origin of these clouds should be a relatively `common' one, and casts doubt on the intermediate black hole scenario.

In this paper we re-examine the supernova scenario. A supernova shock wave in a molecular cloud will be radiative \citep{chevalier_1999,wheeler_et_al_1980} and hence the temperature is expected to be much lower than that of a typical supernova remnant whose shock wave encompasses a comparable ISM mass. This effect significantly suppresses the X-ray emission.

The paper is organized as follows. In section \ref{sec:sn_temp} we explain the absence of the expected thermal emission from a supernova embedded in a molecular cloud. In section \ref{sec:imbh_imp} we present possible issues with the association of intermediate mass black holes with dense molecular clouds with high velocity spreads. In section \ref{sec:conclusion} we summarise and discuss some general implications of these results.

\begin{table*}
\begin{center}
 \begin{tabular}{||cccccccc||} 
 \hline
 Designation & Size [pc] & $\Delta V$ [km/s] & Mass [$M_{\odot}$] & $M_{\rm vir}^1$ [$M_{\odot}$] & $E_k^2$ [erg] & $t_{\rm age}^3$ [yrs] & ref. \\ [0.5ex] 
 \hline
 CO 0.02-0.02 & 3.5 & 100 & $9\cdot 10^4$ & $8.1\times 10^6$ & $1.3\times 10^{51}$ & $7\times 10^4$ & \cite{oka_white_et_al_1999} \\ 
 CO 1.28+0.06 & 8 & 100 & $\approx 4\cdot 10^4$ & $1.9 \times 10^7$ & $6\times 10^{50}$ & $1.6\times 10^5$ & \cite{tokuyama_oka_et_al_2017}\\
 CO 2.88+0.08 & 5 & 80 & $\approx 10^4$ & $7.4\times 10^6$ & $10^{50}$ & $1.2 \times 10^5$ & \cite{tokuyama_oka_et_al_2017}\\
 CO 3.34+0.43 & 8 & 30 & $\approx 4\cdot 10^4$ & $1.7\times 10^6$ & $5\times 10^{49}$ & $5.3\times 10^5$ & \cite{tokuyama_oka_et_al_2017}\\
 CO 0.40-0.22 & 5 & 100 & 4000 & $1.2\times 10^7$ & $6\times 10^{49}$ & $10^5$ & \cite{oka_mizuno_et_al_2016}\\ 
 HCN 0.009-0.044 & 0.37 & 20 & 16 & $3.4\times 10^4$ & $10^{46}$ & $3.7\times 10^4$ & \cite{takekawa_oka_2017}\\
 HCN 0.085-0.094 & 0.33 & 40 & 13 & $1.2 \times 10^5$ & $3\times 10^{46}$ & $1.6 \times 10^4$ & \cite{takekawa_oka_2017}\\ [1ex] 
 \hline
\multicolumn{8}{c}{1 Virial mass of the cloud estimated from size and velocity, $M_{\rm vir}=G \Delta v^2 r$.}\\
\multicolumn{8}{c}{2 Kinetic energy of the cloud assuming a uniform sphere undergoing homologous expansion, $E_k=\frac{3}{10} M \Delta v^2$.}\\
\multicolumn{8}{c}{3 lifetime of cloud estimated from size and velocity, $t_{\rm age}=r/\Delta v$.}\\
 \end{tabular}
 \caption{Parameters of the high velocity compact molecular clouds. Approximate sign in the mass column indicates that a line based mass estimate was not available. In those cases, mass was estimated by multiplying the volume of the cloud with the average density of the line emitting gas ($10^{4.5} cm^{-3}$) times a reasonable filling factor 0.1.} 

 \label{tab:hvcc_params}
 \end{center}
\end{table*}

\section{Supernova Scenario} \label{sec:sn_temp}
\begin{figure}
\includegraphics[width=0.9\columnwidth]{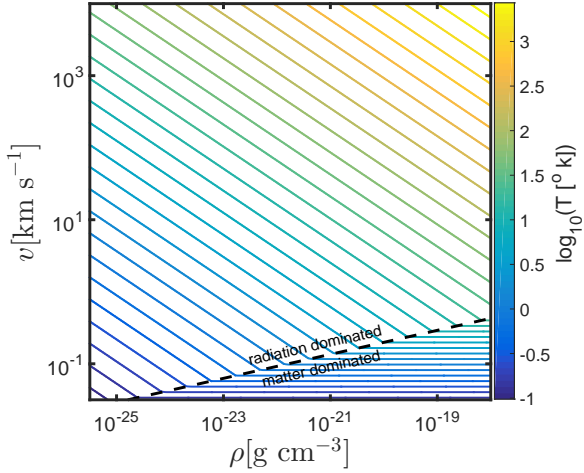}
\caption{Map of the equilibrium temperature as a function of the shock temperature and upstream density.}
\label{fig:tempmap}
\end{figure}

A slow shock wave moving through cold interstellar medium will heat it up to a temperature $k_B T \approx m_p v^2$, where $m_p$ is the proton mass, and $v$ is the velocity of the shock. This is called a matter dominated shock. Higher velocity shocks will be radiation dominated, in which case the temperature is given by $T \approx \left(\rho v^2/a\right)^{1/4}$, where $\rho$ is the mass density and $a$ is the radiation constant \citep{katz_et_al_2010}. In equilibrium, the temperature will be the lower of the two. Figure \ref{fig:tempmap} shows a map of the equilibrium temperature as a function of shock velocity and upstream density. In all astrophysical shocks considered in this work the equilibrium state is radiation dominated. However, some shocks could still emit as though they are matter dominated because they are so dilute that they cannot produce enough photons to reach (the radiation dominated) thermal equilibrium within their lifetime. Such is the case for a typical supernovae remnant. At high temperatures, when matter is completely ionised, the dominant emission mechanism is thermal Bremsstrahlung. The emissivity of this process is given by \citep{Rybicki86} 
\begin{equation}
\varepsilon_{bs} \approx \alpha \frac{m_e c^3}{r_e^4} \left(r_e^3 \frac{\rho}{m_p}\right)^2 \sqrt{\frac{k_B T}{m_e c^2}} \label{eq:bs_emis}
\end{equation}
where $\alpha$ is the fine structure constant, $c$ is the speed of light in vacuum, $m_e$ is the electron mass and $r_e$ is the classical electron radius. 
The emissivity in equation \ref{eq:bs_emis} is roughly accurate for $T>10^9 K$, but underestimates the actual emissivity at lower temperatures due to line emission \citep{sutherland_dopita_1993}.
To reach a Planck thermal equilibrium, a fluid element has to produce enough photons \citep{svensson_1984}. This photon density is given by \citep{nakar_sari_newtonian_breakout_2010}
\begin{equation}
n_{bb} \approx \frac{aT^4}{k_B T}\approx a^{1/4} v^{3/2} \rho^{3/4} / k_B \, .
\end{equation}
where the subscript of $n_{bb}$ stands for black body.
The photon production rate ($\varepsilon_{bs}/k_B T$) decreases with temperature, so it is at its minimum at the matter dominated temperature. The number of photons necessary to reach Planck equilibrium depends on the radiation dominated temperature. Their ratio gives an upper bound on the timescale to reach Planck equilibrium 
\begin{equation}
t_{bb} \approx 2 \cdot 10^{14} \left(\frac{v}{1000 \, \rm km/s}\right)^{5/2} \left(\frac{\rho}{1.6 \cdot 10^{-24} \, \rm gram}\right)^{-5/4} \, \rm {yrs} \, . \label{eq:thermalisation_time}
\end{equation}
The actual timescale is somewhat shorter, because as photons are produced the temperature drops, and the photon production rate increases. Another effect that further accelerates cooling is called a thermal instability. This is a positive feedback loop, where cooling leads to a reduction in the effective adiabatic index, which leads to an increase in the density of the shocked material, which leads to an increase in the cooling rate and so on \citep{blondin_et_al_1998}.

The mass of CO-0.40-0.22 is estimated at 4000 solar masses \citep{oka_mizuno_et_al_2016}. A typical supernova encompasses this much mass towards the end of its life, when the radius is a few tens of parsecs, and its age is a few $10^4$ years \citep{blondin_et_al_1998}. According to equation \ref{eq:thermalisation_time}, the time needed to reach thermal equilibrium is much larger than the age of the supernova, and hence the temperature is matter dominated. At such high temperatures, the shocked ISM is completely ionised, so opacity is solely due to Compton scattering. The optical depth of a typical supernova remnant  with a radius of a few tens of parsecs is
\begin{equation}
\tau \approx R \kappa_T \rho \approx 10^{-4} \frac{R}{\rm 50 \, pc} \frac{\rho}{1.6 \cdot 10^{-24} \, \rm gram/cm^{-3}}
\end{equation}
where $R$ is the radius of the supernova remnant. $\tau \ll 1$ implies that any photon generated will immediately escape the system, and its energy will be lost from the supernova remnant.

In contrast, let us consider a supernova in a much denser environment. The molecular cloud CO-0.40-0.22 is thought to have been much more compact prior to the start of its current expansion. We assume a size of 0.2 pc, adopting the value used in \citep{oka_mizuno_et_al_2016}, although our conclusion will also hold for more compact initial clouds. A supernova shock wave propagating according to a Sedov Taylor solution \citep{sedov1993similarity,taylor1950formation} will cross this distance within about $t=(\rho/E)^{1/2}r^{5/2}=2000$ years (where $E$ is the blast wave energy and $\rho$ is the mass density). The time needed to reach a Planck spectrum is only 200 years. 
The optical depth to Compton scattering at that point is 8. 
Therefore, the shock wave can attain Planck equilibrium, which means most of the energy is in photons and the temperature is 900 K. 
These photons, however, are unable to escape because the optical depth is larger than unity. 

The shock eventually reaches the edge of the cloud, after which the cloud continues to expand at a constant velocity, because the gas around the cloud is much more tenuous \citep{jm_cordes_et_al_1993}. In contrast to regular supernova remnants, where the optical thickness increases with time (but never making it to unity before the onset of radiative cooling), the optical thickness of the freely expanding cloud decreases with time. The cloud continues to cool both adiabatically, and radiatively, since its optical depth to Compton scattering has dropped below unity. However, even after the thermal energy has been exhausted, the kinetic energy remains almost unchanged. By the time the cloud reaches its current size, adiabatic cooling alone will drop the temperature below 90 K. Such low temperature precludes any emission in the X-ray range. Even in lower wavelengths it would be hard to observe, as it is of the same order of magnitude as the background \citep{oka_hasegawa_et_al_1999}.

One possible issue with the supernova scenario is that at the shock front the temperatures are still very high, and this can destroy molecules. These molecules are expected to reform after the transition of the shock front, but this process can take a long time, possible longer than the lifetime of the system. This difficulty could be averted if the shock develops a precursor wave due to photons leaking to the upstream material \citep{draine_mckee_1993}. These precursors can drastically decrease the peak temperature attained as a fluid element traverses the shock front and travels downstream.

The supernova scenario is also supported by a statistical argument. A cloud formation rate of $10^{-5} \rm{yrs}^{-1}$ is consistent with a supernova origin, as these have a much larger rate of $10^{-2} \rm{yrs}^{-1}$ in the Galaxy \citep{diehl_et_al_2006}. Furthermore, massive star forming regions and hence the supernovae that follow them, are known to preferentially occur within such dense molecular clouds \citep{lada_lada_2003,saito_et_al_2007,heyer_et_al_2016}. This is consistent with the observation of ``shell'' features, thought to be cavities blown by multiple supernovae, next to some high velocity clouds \citep{oka_mizuno_et_al_2016}.

\section{Implications of Intermediate Mass Black Holes} \label{sec:imbh_imp}
So far, about 4 such clouds have been detected at distances of a few hundred parsecs from the galactic center \citep{oka_hasegawa_et_al_1999}, one at a distance of 60 pc \citep{oka_mizuno_et_al_2016}, and two within 20 pc \citep{takekawa_oka_2017} (see table table \ref{tab:hvcc_params}). Assuming all such clouds are due to an intermediate mass black hole and assuming that they each correspond to a unique black hole, we can get a lower limit on the number of such black holes. The lifetime of these clouds is of the order of a hundred thousand years, after which they break up completely. Assuming a constant formation rate, this corresponds to a rate of at least $r_c\approx 10^{-4} \rm{yr}^{-1}$. Thus, roughly $10^6$ such clouds have been formed over the Milky way's evolution and the total mass of the black holes within 300 pc of the galactic center would be huge, about $10^{11} M_{\odot}$, which is larger than that of the galactic bulge.

Each intermediate black hole could, in principle,  produce multiple high velocity clouds. This will of course reduce the requirement on the total mass in intermediate black holes. However, this will require a large number of such clouds, and hence a large mass. The net mass of the high velocity clouds needed can be estimated by assuming random, isotropic collisions is
\begin{equation}
M_c \approx \frac{m_c r_o^3 r_c}{N_h b_c^2 v_v} \approx 2.6 \cdot 10^9 \frac{m_c}{10^4 M_{\odot}} \left(\frac{r_o}{300 \, \rm pc}\right)^3 \times
\end{equation}
\begin{equation*}
\times \left(\frac{N_h}{10^2}\right)^{-1} \left(\frac{b}{1 \, \rm pc}\right)^{-2} \left(\frac{r_c}{10^{-4}\, \rm yrs^{-1}}\right)\left(\frac{v_v}{100 \, \rm km/s}\right)^{-1} M_{\odot}
\end{equation*} 
where $m_c$ is the average mass of a high velocity cloud, $r_o$ is the galacto-centric radius to which they were observed, $N_h$ is the number of black holes, $v_v$ is the virial velocity and $b_c$ is the critical impact parameter between a cloud and a black hole below which the cloud will be disrupted. The total mass of these clouds decreases with the number of black holes. For $10^2$ intermediate mass black holes, as suggested by some studies \citep{rashikov_madau_2013}, the total mass in high velocity clouds becomes more than an order of magnitude larger than the total mass enclosed within the central 300 parsec, which is about $10^8 M_{\odot}$.

\begin{figure}
\includegraphics[width=0.9\columnwidth]{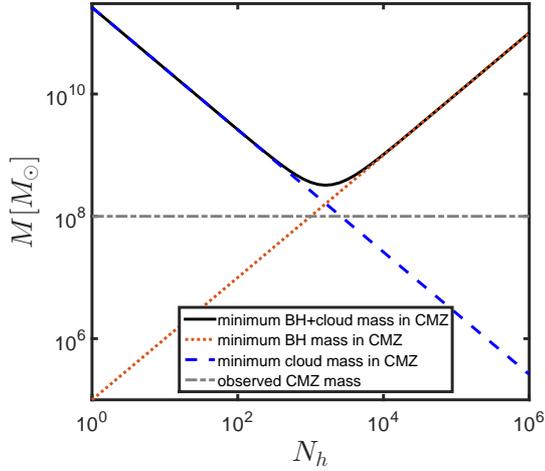}
\caption{Total implied mass in intermediate mass black holes and in high velocity compact clouds, assuming that all high velocity clouds are due to interactions with such black holes. Regardless of the number of black holes, the implied mass is larger than the entire observed mass in the CMZ.}
\label{fig:CMZmass}
\end{figure}

Figure \ref{fig:CMZmass} depicts the total implied mass in intermediate black holes and high velocity clouds within the inner 300 parsecs (an area also known as the central molecular zone or CMZ). As can be seen from the figure, under the assumption that all high velocity clouds are accelerated by black holes, the sum of the clouds and black hole masses is larger than the entire observed mass of the CMZ for any number of intermediate black holes.

Another argument limiting the number of intermediate mass black holes is the rate of tidal disruption events. A black hole will tidally disrupt stars within their loss cone \citep{rees_1988}. The cross section for such an event is given by
\begin{equation}
\sigma_{lc} \approx \frac{G M_h}{v_v^2} R_s \left(\frac{M_h}{m_s} \right)^{1/3}
\end{equation}
where $G$ is the universal constant of gravity, $M_h$ is the mass of the intermediate mass black hole, $R_s$ is the stellar radius and $m_s$ is the mass of the star. The time scale for tidal disruption of stars by a single intermediate mass black hole is
\begin{equation}
t_{lc} \approx 2 \cdot 10^9  \left(\frac{n_s}{10^2 \, \rm pc^{-3}}\right)^{-1} \frac{v_v}{10^2 \, \rm km/s} \times
\end{equation}
\begin{equation*}
\times \left(\frac{M_h}{10^5 M_{\odot}}\right)^{-4/3} \left(\frac{R_s}{R_{\odot}}\right)^{-1} \left(\frac{m_s}{M_{\odot}}\right)^{1/3}\, \rm year
\end{equation*}
where $n_s$ is the number density of stars. When the number of intermediate mass black holes exceeds $10^5$, then the rate of tidal disruption events will exceed the the theoretical rate expected due to the supermassive black hole at the galactic centre (once per $10^4$ years). The rate inferred from observations is even lower \citep{stone_metzger_2016}, which sets an even more stringent upper limit on the number of intermediate mass black holes.

\section{Conclusions} \label{sec:conclusion}
In this paper we discuss the recent suggestion that a compact cloud with high velocity spread serves as evidence for an intermediate mass black hole. We propose a less exotic scenario, where the same observational features can be reproduced by a supernova. A supernova has enough energy to endow the molecular cloud with its current velocity spread. Since such a supernova exploded in a very dense environment, photon production is much faster, compared with a typical supernova remnant. As more photons are generated, the same amount of energy divides between more particles and the temperature drops. We show that contrary to a typical supernova, material shocked by a supernova in a dense environment can attain Planck equilibrium on a time scale shorter than the dynamical time. We demonstrated this for thermal Bremsstrahlung. Other emission mechanisms such as atomic and molecular lines can accelerate this process even further. 

In addition, we presented two arguments against the association of these clouds with black holes. Due to the short lifetimes of these clouds relative to the age of the galaxy, we estimate that there have been many more such clouds in the past, which have since dispersed. We showed that this implies that the total mass of molecular clouds and intermediate mass black holes exceeds the mass of the volume in which they were observed. The second argument is that these black holes may cause too many tidal disruption events, in excess of the rate inferred from observations. This problem is further exacerbated by the fact that intermediate mass black holes can disrupt compact object like white dwarfs, which bigger holes cannot \citep{rosswog_et_al_2009}.

Another feature associated with CO 0.40-0.22 is a continuum millimeter wavelength source \citep{oka_mizuno_et_al_2016}. The spectrum is sampled at two frequencies, 231 and 266 GHZ, with fluxes 8.4 and 9.9 mJy, respectively. This emission is thought to be blackbody radiation from an object with a temperature of 9 K. We can conceive of two possible explanations for this emission. One option is a protostar, which can reach a similar luminosity and temperature \citep{larson_1969}. Another option is a neutron star. Other neutron stars in the galactic centre produce similar fluxes \citep{schnizteler_et_al_2016}. A neutron star's flux in this frequency typically decreases with frequency, which is inconsistent with the observations. This inconsistency can be due to fluctuations that exceed the error margin, as can be seen in the observations of the Crab pulsar spectrum \citep{arendt_et_al_2011}. Hopefully, future observations at different frequencies will shed light on the nature of the continuum source.

\section*{Acknowledgements}

AY would like to thank Chris Matzner, Chris Thompson and Patrick Breysse for useful discussions. PB would like to thank Joe Silk for helpful suggestions and comments.




\bibliographystyle{mnras}
\bibliography{almogyalinewich} 








\bsp	
\label{lastpage}
\end{document}